\documentclass[12pt]{article}

\usepackage{amssymb}
\usepackage{citesort}
\usepackage{amsmath}
\usepackage{url}

\def\la{\langle}
\def\ra{\rangle}
\def\pa{\partial}

\def\hb{\hbar}

\def\ii{\textrm i}

\newcommand{\db}{de$\,$Broglie}

\newcommand{\dbb}{de$\,$Broglie-Bohm}

\begin{document}
\vspace*{1.0cm}
\noindent
{\bf
{\large
\begin{center}
On Epstein's trajectory model of non-relativistic quantum mechanics
\end{center}
}
}

\vspace*{.5cm}
\begin{center}
Ward Struyve{\footnote{Postdoctoral Fellow FWO.}}
\end{center}

\begin{center}
Institute of Theoretical Physics, K.U.Leuven,\\
Celestijnenlaan 200D, B--3001 Leuven, Belgium.{\footnote{Corresponding address.}}\\
Institute of Philosophy, K.U.Leuven,\\
Kardinaal Mercierplein 2, B--3000 Leuven, Belgium.\\
E--mail: Ward.Struyve@fys.kuleuven.be.
\end{center}

\begin{abstract}
\noindent
In 1952 Bohm presented a theory about non-relativistic point-particles that move deterministically along trajectories and showed how it reproduces the predictions of standard quantum theory. This theory was actually presented before by \db\ in 1926, but Bohm's particular formulation of the theory inspired Epstein to come up with a different trajectory model. The aim of this paper is to examine the empirical predictions of this model. It is found that the trajectories in this model are in general very different from those in the \db-Bohm theory. In certain cases they even seem bizarre and rather unphysical. Nevertheless, it is argued that the model seems to reproduce the predictions of standard quantum theory (just as the \db-Bohm theory).
\end{abstract}

\renewcommand{\baselinestretch}{1.1}
\bibliographystyle{unsrt}

\section{Introduction}
In the \dbb\ theory, discovered by \db\ \cite{debroglie28} and rediscovered (in a different form) by Bohm \cite{bohm52a,bohm52b}, an individual closed system is described by its wavefunction $\psi(x,t)$ on configuration space, with $x=({\bf x}_1,\dots,{\bf x}_N)$, which satisfies the non-relativistic Schr\"odinger equation
\begin{equation}
\ii \hb \frac{\pa \psi(x,t)}{\pa t} = - \sum^N_{k=1} \frac{\hb^2}{2m_k}\nabla^2_{k}\psi(x,t) + V(x) \psi(x,t) \,,
\label{1}
\end{equation}
and by particle positions ${\bf x}_1(t),\dots,{\bf x}_N(t)$, whose possible trajectories are solutions of the {\em guidance equations}
\begin{equation}
\frac{d {\bf x}_k }{dt} = \frac{ 1}{m_k}{\boldsymbol \nabla}_k S \,,
\label{2}
\end{equation}
where $S$ is the phase of the wavefunction, that is, $\psi = |\psi| \exp(\ii S/\hb)$.

An important property of the guidance equations is that they preserve the distribution $|\psi|^2$. That is, if the distribution of particles is given by $|\psi(x,t_0)|^2$ at a certain time $t_0$, then the distribution is given by $|\psi(x,t)|^2$ at other times $t$. This property is called {\em equivariance} \cite{durr92} (and $|\psi|^2$ is actually the only distribution that is a suitably local functional of $\psi$ that satisfies it \cite{goldstein07}). Equivariance essentially follows from the fact that the distribution $|\psi(x,t)|^2$ satisfies the continuity equation
\begin{equation}
\frac{\pa |\psi|^2}{\pa t} +  \sum^N_{k=1} {\boldsymbol \nabla}_k  \cdot \left( \frac{ {\boldsymbol \nabla}_k S}{m_k} |\psi|^2 \right) = 0 \,,
\label{3}
\end{equation}
just as any distribution $\rho(x,t)$ that is transported along the \db-Bohm trajectories. The distribution $|\psi|^2$ plays the role of an equilibrium distribution (see for example \cite{bohm53b,valentini91a,durr92}) and is called the {\em quantum equilibrium distribution}. Given the quantum equilibrium distribution and the fact that measurement results are generally ultimately recorded in positions of macroscopic pointers, like instrument needles, computer outprint, etc., it almost follows immediately that the \dbb\ theory reproduces the standard quantum mechanical predictions (see for example \cite{bohm52b,bohm93,holland93b,durr033,durr09}).

Instead of following the above presentation of the theory, which is along the lines of that of \db, Bohm actually presented the theory in a Newtonian form. Bohm observed that by using the polar decomposition of the wavefunction, i.e.\ $\psi = |\psi| \exp(\ii S/\hb)$, the Schr\"odinger equation yields an equation that is similar to the Hamilton-Jacobi equation of Newtonian mechanics. This inspired Bohm to adopt a Newtonian type equation with an extra $\psi$-dependent potential as the basic equation of motion, thereby regarding the guidance equations \eqref{2} as constraints on the possible momenta.

It was this alternative formulation of Bohm that inspired Epstein to come up with an alternative trajectory model \cite{epstein52,epstein53}. Epstein found this model by considering the wavefunction in momentum space rather than in configuration space and by following an analysis similar to that of Bohm. 

The aim of this paper is to examine the empirical predictions of Epstein's theory. To our knowledge this has not been done before. It is found that localized macroscopic systems, such as pointers in measurement situations, seem to be located at locations predicted by standard quantum theory and this with the same probabilities as given by standard quantum theory. This indicates that Epstein's trajectory theory seems to reproduce the standard quantum predictions, just as the theory of \db\ and Bohm.

\section{Epstein's trajectory model}\label{etm}
Just as Bohm, Epstein derived his model by considering the formal analogy with classical Hamilton-Jacobi theory. In presenting Epstein's model we will not consider this analogy, but instead restrict ourselves to give just the basic equations of motion (just as \eqref{1} and \eqref{2} can be regarded as the basic equations of motion in the \dbb\ theory). 

\subsection{Equations of motion}
In Epstein's trajectory model the wavefunction is written in the momentum representation, by taking the Fourier transform of the wavefunction in the position representation:
\begin{equation}
{\widetilde \psi} (p,t) \equiv {\mathcal{F}}_p(\psi(x,t)) \equiv  (2\pi\hb)^{-\frac{3N}{2}}  \int d x e^{-\ii x \cdot p/\hb} \psi(x,t) \,.
\label{4}
\end{equation}
Application of the Fourier transform to both sides of the non-relativistic Schr\"odinger equation \eqref{1} yields
\begin{equation}
\ii \hb \frac{\pa }{\pa t} {\widetilde \psi} (p,t) = \sum^N_{k=1} \frac{p^2_k }{2m_k}{\widetilde \psi} (p,t) + {\mathcal{F}}_p(V(x)\psi(x,t))\,.
\label{5}
\end{equation}
This implies
\begin{equation}
 \frac{\pa |{\widetilde \psi}(p,t)|^2 }{\pa t} + I^\psi(p,t) = 0\,,
\label{6}
\end{equation}
where
\begin{equation}
I^\psi(p,t) \equiv \frac{2}{\hb} {\textrm{Re}} \left( \ii{\widetilde \psi}^* (p,t) {\mathcal{F}}_p(V(x)\psi(x,t)) \right)\,.
\label{7}
\end{equation}
For Epstein's model the quantity $I^\psi$ needs to be written as a divergence $\sum_k {\widetilde {\boldsymbol \nabla}}_k \cdot {\bf j}^\psi_k$, where ${\widetilde {\boldsymbol \nabla}}_k = (\pa / \pa p_{kx},\pa / \pa p_{ky},\pa / \pa p_{kz})$ and ${\bf p}_k =(p_{kx},p_{ky},p_{kz}) $, in such a way that the above equation \eqref{6} can be interpreted as a continuity equation for the momentum density $|{\widetilde \psi}|^2$ with currents ${\bf j}^\psi_k$ (in particular the currents should die off sufficiently fast for $|p| \to \infty$ so that there is no flux at infinity). The possibility of introducing such currents depends of course on the form of the potential. This is discussed below.

Given currents ${\bf j}^\psi_k$, the first step in Epstein's model is to introduce the variable{\footnote{One could call these variables momenta. However, it should be kept in mind that the relation ${\bf p}_k = m {\dot {\bf x}}_k$, with ${\bf x}_k$ the $k$-th particle position as defined in \eqref{9}, in general does not hold. The exact relation to the velocity could be found by differentiating the expression in \eqref{9} with respect to time.}} $p=({\bf p}_1,\dots,{\bf p}_N)$ whose time evolution is determined by the differential equations
\begin{equation}
\frac{d {\bf p}_k }{dt} = \frac{ {\bf j}^\psi_k }{|{\widetilde \psi}|^2} \,.
\label{8}
\end{equation}
The next step is to introduce particle positions ${\bf x}_1,\dots,{\bf x}_N$, whose trajectories are defined as 
\begin{equation}
{\bf x}_k(t) \equiv {\textrm{Re}}\left( \frac{{\widetilde \psi}^* (p,t) (\ii \hb {\widetilde {\boldsymbol \nabla}}_k ) {\widetilde \psi} (p,t)}{|{\widetilde \psi} (p,t)|^2} \right) \Bigg|_{p=p(t)} = - {\widetilde {\boldsymbol \nabla}}_k {\widetilde S} (p,t) \bigg|_{p=p(t)} 
\label{9}
\end{equation}
where the polar decomposition ${\widetilde \psi} = |{\widetilde \psi}| \exp(\ii {\widetilde S}/\hb )$ is used and where the expressions on the right hand side are evaluated for an actual trajectory of the variable $p$.{\footnote{These expressions correspond to Holland's local expectation values \cite{holland93b} for the position operators ${\widehat {\bf x}}_k$ in the momentum representation, where ${\widehat {\bf x}}_k \to \ii \hb {\widetilde {\boldsymbol \nabla}}_k $, evaluated for the actual value of $p$.}} 

So in Epstein's trajectory model systems are described by the triplet $(x,p,\psi)$ (where the particle positions constitute the primitive ontology \cite{durr92,allori06}). The wavefunction $\psi$ determines the time evolution of the variable $p$ and the wavefunction together with the variable $p$ determine the configuration $x$.

Let us now return to the construction of the currents ${\bf j}^\psi_k$. The construction of such currents was discussed in detail in \cite{struyve09a}. If the potential is given by a differential operator in momentum space, that is, if there exist functions $V_n(p)$ such that 
\begin{equation}
{\mathcal{F}}_p(V(x)\psi(x,t)) = \sum_{n \ge 0} V_n(p) D^n {\widetilde \psi}(p,t)\,,
\label{10}
\end{equation}
where $n=(n_{1x},n_{1y},n_{1z},\dots,n_{Nx},n_{Ny},n_{Nz}) \in {\mathbb N}^{3N}$ and $D^n = (\pa^{n_{1x}} / \pa p_{1x}^{n_{1x}}, \dots, \pa^{n_{Nz}} / \pa p_{1z}^{n_{Nz}})$, then currents can be found of the form 
\begin{equation}
{\bf j}^\psi_k(p,t) = \sum_{n,m \ge 0} {\bf J}_{k,nm}(p) D^n {\widetilde \psi} (p,t)D^m {\widetilde \psi}^* (p,t) \,, 
\label{11}
\end{equation}
where ${\bf J}_{k,nm}$ are certain functions that depend on the $V_n$. 

In the case the potential is not given by a differential operator in momentum space, for example if it contains a Coulomb potential, then one could solve the Poisson equation ${\widetilde \nabla}^2 F = I^\psi$, where ${\widetilde \nabla}^2  = \sum_k {\widetilde \nabla}^2_k$ is the $3N$-dimensional Laplacian, and define ${\bf j}^\psi_k = {\widetilde {\boldsymbol \nabla}}_k F$. A solution is given by $F={\widetilde \nabla}^{-2} I^\psi$ \cite{struyve09a} (given that it is well-defined), in which case
\begin{equation}
{\bf j}^\psi_k(p,t) ={\widetilde {\boldsymbol \nabla}}_k  \frac{1}{{\widetilde \nabla}^{2}} I^\psi(p,t)  \,. 
\label{12}
\end{equation}
This method of constructing a current was suggested by Epstein himself \cite{epstein53} in response to a worry raised by Bohm \cite{bohm53a} concerning the treatment of certain potentials such as the Coulomb potential. In the following we just assume that there exists a suitable choice of current, so that the dynamics \eqref{8} for the variable $p$ can be introduced.

\subsection{Ensemble distribution}\label{qed}
It is assumed that the distribution of the variable $p$ over an ensemble of systems all described by the same wavefunction ${\widetilde \psi}$ is given by $|{\widetilde \psi}|^2$. This distribution is equivariant with respect to the time evolution of the variable $p$.

The implied distribution for the configuration $x$ is then given by 
\begin{equation}
\rho^\psi(x,t) = \int dp |{\widetilde \psi}(p,t)|^2 \delta[x + {\widetilde  \nabla}  {\widetilde S} (p,t)]\,,
\label{13}
\end{equation}
where we defined
\begin{equation}
\delta[x + {\widetilde  \nabla}  {\widetilde S} (p,t)] = \prod^N_{k=1} \delta[ {\bf x}_k + {\widetilde {\boldsymbol \nabla}}_k {\widetilde S} (p,t) ] \,.
\label{14}
\end{equation}
The distribution $\rho^\psi$ could be regarded as an equilibrium distribution, analogous to the quantum equilibrium distribution $|\psi|^2$ in the theory of \db\ and Bohm. However, the distribution $\rho^\psi$ is in general very different from $|\psi|^2$ (this is illustrated in an example below). Only the latter equals the quantum mechanical position distribution.

While the position distribution $\rho^\psi$ is in general different from the quantum mechanical one, the expectation value for position equals the quantum mechanical one.{\footnote{This is a defining property of Holland's local expectation value.}} That is, if we write ${\bf x}_k=(x_k,y_k,z_k)$, then 
\begin{equation}
\la x_k \ra = \int dx  \rho^\psi(x)x_k = - \int dp |{\widetilde \psi}(p)|^2 \frac{\pa {\widetilde S}}{ \pa p_{kx}} (p)= \int dp {\widetilde \psi}^*(p) \left( \ii \hb \frac{\pa }{ \pa p_{kx}}\right) {\widetilde \psi}(p)  = \la {\widehat x}_k \ra 
\label{15}
\end{equation}
and similarly for the other Cartesian coordinates. We also have that the standard deviation satisfies the inequality $\Delta x_k \leqslant \Delta {\widehat x}_k$, and similarly for the other Cartesian coordinates. This follows from 
\begin{equation}
\la x^2_k \ra = \la {\widehat x}^2_k \ra - \hb^2 \int dp \left( \frac{\pa}{\pa p_{kx}} |{\widetilde \psi}(p)| \right)^2 \leqslant \la {\widehat x}^2_k \ra \,.
\label{16}
\end{equation}

\subsection{Example}
Consider a free particle as an example. Its wavefunction in momentum space is given by
\begin{equation}
{\widetilde \psi}({\bf p},t) = f({\bf p}) e^{-\ii p^2 t / 2m\hb}
\label{17}
\end{equation}
where $f$ is an arbitrary (square integrable) function. Suppose further that $f$ is real. Since the potential is zero the evolution equation \eqref{8} reduces to $d{\bf p}/dt =0$, so that the possible trajectories are given by 
\begin{equation}
{\bf x}(t) = \frac{ {\bf p} t }{m}\,,
\label{18}
\end{equation}
where ${\bf p}$ is a constant vector, distributed according to $|{\widetilde \psi}({\bf p}, t)|^2=f({\bf p})^2$ over an ensemble. Note that all the trajectories pass through the origin at $t=0$. So clearly the initial position distribution $\rho^\psi({\bf x},0)=\delta({\bf x})$ is different from the quantum mechanical position distribution $|\psi({\bf x},0)|^2$. Also the trajectories are significantly different from the trajectories in the theory of \db\ and Bohm. For example the trajectories in the latter model never cross in configuration space, while our example illustrates that the trajectories in Epstein's model are clearly allowed to do that. 

The particular position distribution in Epstein's model seems to lead to rather bizarre consequences. Consider again a free particle, with wavefunction in the superposition
\begin{equation}
\phi({\bf x},t)=\frac{1}{{\sqrt 2}} (\psi({\bf x} - {\bf a},t) + \psi({\bf x} + {\bf a},t)) \,,
\label{19}
\end{equation}
where $\psi({\bf x},t)$ is a wavefunction localized near the origin at time $t=0$ and where ${\bf a}$ is some constant vector. In the momentum representation this becomes
\begin{equation}
{\widetilde \phi}({\bf p},t) = {\sqrt 2} \cos({\bf a} \cdot{\bf p} /\hb) {\widetilde \psi}({\bf p},t)  \,.
\label{20}
\end{equation}
With ${\widetilde \psi}$ of the form \eqref{17} with $f$ real, the particle distribution for an ensemble is $\rho^\psi({\bf x},0)=\delta({\bf x})$. While we would have expected that the particles are typically located around ${\bf a}$ or $-{\bf a}$ at time $t=0$, they are located at the origin, independently of the value of ${\bf a}$. For example, the wavefunction of the particle in the double slit experiment is of the form \eqref{19} (after it has passed the slits) and the above analysis implies that the actual particle position would initially come from the midpoint between the slits. Hence, Epstein's model seems to yield a bizarre and rather unphysical picture for isolated particles in certain cases. Nevertheless we will argue in the next section that the model actually seems capable of reproducing the predictions of standard quantum theory, because it can account for positions of macroscopic systems.

\section{Reproducing the predictions of standard quantum theory}
\subsection{Macroscopic systems}\label{lmo}
Consider a wavefunction $\psi(x)$ that represents a macroscopic system, like for example an instrument needle or a cat. This wavefunction has most of its support on configurations that on the macroscopic level give the image of the macroscopic system. That is, those configurations all correspond to the same macrostate of the system. Let us call this set of configurations $B$. In the theory of \db\ and Bohm, where the configurations are distributed according to $|\psi(x)|^2$ over an ensemble, the configurations are typically in $B$. So the particles typically yield the image of the macroscopic system. This is true also for Epstein's theory: configurations are typically within $B$ because the distribution $\rho^\psi$ satisfies $\la x_k \ra = \la {\widehat x}_k \ra$ and $\Delta x_k \leqslant \Delta {\widehat x}_k$, cf.\ Section \ref{qed}, and because $B$ can be assumed to have a suitably simple form.

Consider now wavefunctions that are in a superposition of macroscopically different states of the macroscopic system. First assume that the wavefunction $\psi(x)$ represents a macroscopic system that is localized near the origin of our reference frame in physical 3-space and call $B$ again the set of configurations that on the macroscopic level yield the image of the macroscopic system. The wavefunction $\psi_a(x)=\psi(x-a)$, with $a=({\bf a}, \dots, {\bf a})$, will represent the same system located near ${\bf a}$. Consider now the superposition 
\begin{equation}
\phi(x)=\frac{N}{{\sqrt 2}} (\psi_a(x) + \psi_{-a}(x)) \,,
\label{49}
\end{equation}
where $N$ is a normalization factor. If $|{\bf a}|$ is sufficiently large, the wavefunctions $\psi_a$ and $\psi_{-a}$ will only have negligible overlap, so that $N \approx 1$. This is definitely the case if $|{\bf a}|$ is much bigger then the size of the macroscopic system. In the theory of \db\ and Bohm the fact that those wavefunctions have only negligible overlap implies that the configuration will typically display the macroscopic system localized at either ${\bf a}$ or $-{\bf a}$. This is how the theory deals with Schr\"odinger's cat paradox. If the wavefunction of a system is in a superposition of macroscopically distinct states, then the particle configuration will typically display only one of those macroscopic states.

In Epstein's theory the situation is a bit different. Let us analyze what the typical particle configuration is for the state $\phi$. First of all, the Fourier transform of $\phi$ is given by
\begin{equation}
{\widetilde \phi}(p) = {\sqrt 2}N \cos(a \cdot p/\hb) {\widetilde \psi}(p)  \,,
\label{50}
\end{equation}
where ${\widetilde \psi}$ is the Fourier transform of $\psi$. Hence the position distribution is given by 
\begin{equation}
\rho^\phi (x) = 2 N^2 \int dp \cos^2(a \cdot p/\hb) |{\widetilde \psi}(p)|^2  \delta[x + {\widetilde  \nabla}  {\widetilde S} (p)] \,, 
\label{51}
\end{equation}
where ${\widetilde S}$ is the phase of ${\widetilde \psi}$. Since 
\begin{equation}
\rho^\phi (x) \leqslant 2N^2 \int dp |{\widetilde \psi}(p)|^2  \delta[x + {\widetilde  \nabla}  {\widetilde S} (p)]   = 2 N^2 \rho^\psi (x)
\label{52}
\end{equation}
and since $\rho^\psi$ has negligible support outside $B$ (and $N \approx 1$), also $\rho^\phi$ must have negligible support outside $B$. Hence the position configuration typically does not display the macroscopic systems localized near ${\bf a}$ or $-{\bf a}$, but rather near the origin.

This feature seems to suggest that Epstein's model will fail to reproduce the predictions of standard quantum theory. However this is not so. The reason is that we did not include the description of the environment in our analysis. The introduction of the environment drastically changes the description. To see this, consider a general superposition
\begin{equation}
\phi'(x,x_e) = c_1 \psi_1(x) \chi_1(x_e)  + c_2 \psi_{2}(x) \chi_{2}(x_e) \,, 
\label{53}
\end{equation}
where the $c_i$, $i=1,2$, are constants, where the normalized states $\psi_i$ are macroscopically different states of a localized macroscopic system (for example $\psi_1 = \psi_a$ and $\psi_2=\psi_{-a}$) and where the normalized states $\chi_i$ represent that part of the environment that is in direct interaction with the macroscopic system, like for example the air molecules that scatter off the system. This implies that the Fourier transforms ${\widetilde \chi}_{1}$ and ${\widetilde \chi}_{2}$ have negligible overlap in momentum space. It is namely sufficient that there is one air molecule in the environment that scatters off the macroscopic system and obtains a different momentum as a result. In a realistic situation there will be many molecules that obtain a different momentum. As a result, the momentum distribution is approximately given by 
\begin{equation}
|\phi'(p,p_e)|^2 \approx   |c_1|^2  |{\widetilde \psi}_1(p)|^2|{\widetilde \chi}_{1}(p_e)|^2 + |c_2|^2 |{\widetilde \psi}_2(p)|^2|{\widetilde \chi}_{2}(p_e)|^2\,, 
\label{54}
\end{equation}
with $|c_1|^2 + |c_2|^2 \approx 1$ (when the $\chi_i$ are completely non-overlapping these relations hold exactly). The corresponding position distribution is then given by
\begin{align}
\rho^{\phi'} (x,x_e) &\approx   \int dp dp_e 
\bigg(  |c_1|^2 |{\widetilde \psi}_1(p)|^2 \delta[x + {\widetilde  \nabla}  {\widetilde S}_1 (p) ]| {\widetilde \chi}_{1}(p_e) |^2 \delta[x_e + {\widetilde  \nabla}_e  {\widetilde S}_{e,1} (p_e)] \\
& \qquad + |c_2|^2 |{\widetilde \psi}_2(p)|^2 \delta[x + {\widetilde  \nabla}  {\widetilde S}_2 (p)] | {\widetilde \chi}_{2}(p_e) |^2 \delta[x_e + {\widetilde  \nabla}_e  {\widetilde S}_{e,2} (p_e)] \bigg)  \\
&=   |c_1|^2 \rho^{\psi_1} (x)  \rho^{\chi_1} (x_e) + |c_2|^2 \rho^{\psi_2} (x)  \rho^{\chi_2} (x_e)  \,,
\label{55}
\end{align}
where ${\widetilde S}_{e,i}$ are the phases of ${\widetilde \chi}_{i}$, $i=1,2$ and ${\widetilde  \nabla}_e$ represents the derivative with respect to $p_e$. Since the densities $\rho^{\psi_1}$ and $\rho^{\psi_2}$ have approximately different supports (they have there support approximately concentrated on different macrostates), the configuration of the macroscopic system will typically be in the macrostate corresponding to either $\psi_1$ or $\psi_2$ and this with probabilities respectively given by $|c_1|^2$ and $|c_2|^2$, which are the same probabilities as in standard quantum theory. For example, turning back to the particular superposition \eqref{49}, the above analysis shows that when the environment is taken into account, the configurations will typically display the macroscopic system localized either near ${\bf a}$ or $-{\bf a}$, with equal probability. 

Note that in many cases, the states ${\widetilde \psi}_1$ and ${\widetilde \psi}_2$ are themselves non-overlapping. It is namely sufficient that one particle of the system has a different momentum in ${\widetilde \psi}_1$ compared to ${\widetilde \psi}_2$ (this will for example definitely be the case for a superposition of a live and dead cat). In such a case there is actually no need to invoke the environment to establish that the particle configuration will typically display one of the macroscopic states.

\subsection{Collapse of the wavefunction}
In the model of Epstein, the wavefunction evolves according to Schr\"odinger's equation at all times. It never collapses. However, under certain circumstances the wavefunction can undergo an {\em effective collapse}. Effective collapse means the following. Suppose the wavefunction is given by a superposition $\psi=\psi_1 + \psi_2$. Then it can happen that the dynamics of the actual configuration depends only on the wavefunction $\psi_1$ or $\psi_2$ from some time onwards. Since the other wavefunction then does not play any further role in the dynamics of the configuration, it can be removed from the description of the evolution of the configuration. This is called an effective collapse.{\footnote{In the \dbb\ theory, the wavefunction of a subsystem of the universe can be defined as the {\em conditional wavefunction} \cite{durr92} (which is constructed from the universal wavefunction and the actual positions of the particles not belonging to the subsystem). This wavefunction actually undergoes what could be called an actual collapse \cite{durr92}. In the context of Epstein's theory, the wavefunction of a subsystem could be introduced in a similar way (using the universal wavefunction in momentum space and the actual $p$ values not belonging to the subsystem). However, we will not pursue this further here.}}
 
The circumstances under which there is effective collapse depend on the form of the dynamics. In Section \ref{etm}, two possible types of dynamics were considered for the variable $p$. For a potential that is given by a differential operator in momentum space, the current \eqref{11} could be chosen for the equation of motion of the variable $p$. For a potential that is not of such a form the current \eqref{12} could be chosen (given that the expression is well-defined). 

In the case the potential is a differential operator, there is effective collapse if the wavefunction is given by a superposition 
\begin{equation}
{\widetilde \psi}(p,t) = {\widetilde \psi}_1(p,t) + {\widetilde \psi}_2(p,t)\,, 
\label{60}
\end{equation}
where ${\widetilde \psi}_1$ and ${\widetilde \psi}_2$ have negligible overlap in momentum space from some time $t_0$ onwards. This is because the current \eqref{11} then has a similar decomposition:
\begin{equation}
j^\psi(p,t) \approx  j^{\psi_1}(p,t) + j^{\psi_2}(p,t)\,,
\label{61}
\end{equation}
from time $t_0$ onwards. This implies that if the momentum variable $p$ is within the support of ${\widetilde \psi}_l$, $l=1,2$, at time $t_0$, then its future evolution will be determined only by $\psi_l$, through the equation of motion $dp/dt=j^{\psi_l} / |{\widetilde \psi}_l|^2$. In addition we have that 
\begin{equation}
{\bf x}_k(t) \approx  - {\widetilde {\boldsymbol \nabla}}_k {\widetilde S}_l (p,t) \big|_{p=p(t)} 
\label{62}
\end{equation}
with ${\widetilde S}_l$ the phase of ${\widetilde \psi}_l$. Hence the evolution of the particles only depends on the wavefunction ${\widetilde \psi}_l$, so that the other packet in the superposition can be removed from the description of the evolution of the particles. Since the variable $p$ is distributed according to $|{\widetilde \psi}|^2$ over ensembles, it is easily verified that the probability with which an effective collapse to $\psi_l$ occurs is the same as the probability standard quantum theory would assign for an actual collapse of the wavefunction $\psi$ to $\psi_l$.

In the case the potential is not a differential operator in momentum space and the current \eqref{12} is used in the equations of motion for the variable $p$, it is not so clear whether the above conditions on the state \eqref{60} are sufficient to guarantee effective collapse. If, in addition to having negligible overlap, the two wavefunctions ${\widetilde \psi}_l$ both satisfy the Schr\"odinger equation (in the case the potential is given by a differential operator the latter automatically follows from the non-overlap), then $I^{\psi} \approx I^{\psi_1} + I^{\psi_2}$. From the equations $\pa |{\widetilde \psi}_l|^2 / \pa t + I^{\psi_l} = 0$ it follows also that the $I^{\psi_l}$ have approximately the same support as $\psi_l$ (provided $\psi_l$ does not fluctuate too fast outside the region where most of its support is concentrated). As such the functions $I^{\psi_1}$ and $I^{\psi_2}$ will have negligible overlap. However the currents \eqref{12} are non-local in momentum space so that the latter feature does not immediately imply that the components $j^{\psi_1}$ and $j^{\psi_2}$ in the current $j^\psi \approx  j^{\psi_1} + j^{\psi_2}$ are non-overlapping. While the currents $j^{\psi_l}$ die off outside the region where most of the support of $I^{\psi_l}$ is concentrated (the currents $j^{\psi_l}$ relate to $I^{\psi_l}$ in the same way the electric field relates to the charge density by which it is generated), a more detailed analysis would be required to find the rate at which this happens. While equation \eqref{62} holds regardless of this, the configuration will depend on both terms of the wavefunction through the actual value of $p$, if the time evolution of the latter depends on both terms.

\subsection{Measurement situation}
Consider now the description of an ideal quantum measurement in terms of Epstein's theory. Before the measurement it can be assumed that the wavefunction $\psi$ is given by a product wavefunction
\begin{equation}
\psi = \psi_s \psi_m \psi_e \,,
\label{70}
\end{equation}
where the wavefunction $\psi_s$ represents the microscopic system under observation, where $\psi_m$ represents the measurement device which includes some macroscopic pointer, for example an instrument needle, and where $\psi_e$ represents that part of the environment that is in direct interaction with the measurement device. During the measurement the wavefunction will evolve into a superposition, i.e.
\begin{equation}
\psi \to \sum_l c_l \psi_l = \sum_l c_l \psi_{s,l} \psi_{m,l} \psi_{e,l} \,,
\label{71}
\end{equation}
where the wavefunctions $\psi_{s,l}$ represent the different eigenstates of the quantum operator that is being measured, where the $\psi_{m,l}$ represent the corresponding states of the measurement device, including the macroscopic pointer indicating the measurement outcome, where the $\psi_{e,l}$ represent the different associated states of the environment, and where the $c_l$ are constants.

As explained in Section \ref{lmo}, it is guaranteed that the different packets ${\widetilde \psi}_l={\widetilde \psi}_{s,l}{\widetilde \psi}_{m,l}{\widetilde \psi}_{e,l}$ have negligible overlap in momentum space, since at least the packets ${\widetilde \psi}_{e,l}$ have negligible overlap. And as a result, the configuration $x_m$ of the measurement device will typically display one of the possible states of the macroscopic pointer, thereby displaying the outcome of the measurement. The possible results are the same as those expected according to standard quantum theory and are obtained with the same probabilities as predicted by standard quantum theory. As such the standard quantum predictions for measurements are reproduced by Epstein's theory. 

In the case the potential is given by a differential operator in momentum space, the non-overlap of the packets ${\widetilde \psi}_l$ guarantees that there will be an effective collapse to one of these packets with the same probability as in standard quantum theory. In this way the collapse postulate of standard quantum theory is recovered. In the case the potential is not given by a differential operator, it is unclear whether there is an effective collapse. If there is no effective collapse it might be that the actual configuration moves (very fast) between the configurations that display the various macroscopic states. While this feature would be empirically unverifiable because the configurations of the system, the measurement device and the observer's memory would maintain their correlation, it seems unacceptable for a physical theory. In any case, even if there is no effective collapse, the model reproduces the predictions of standard quantum theory because, again, the possible measurement results are the same as those in standard quantum theory and they are obtained with the same probabilities.

\section{Conclusion}
While Epstein's trajectory model is very different from that of \db\ and Bohm, it seems to agree with the latter on the empirical level. Examples of other such trajectory models have been given before \cite{deotto98,holland98,goldstein042}. Epstein's model differs at least in one aspect from those theories, namely that the position distribution is in general very different from $|\psi|^2$. An example of another theory which shares this property is that of de Polavieja \cite{depolavieja96a,depolavieja96b}, which is developed using the Schr\"odinger equation in a phase space representation. The extent of agreement of the latter theory with standard quantum theory still needs to be investigated. 

Similar models could be constructed starting from representations intermediate to the position and momentum representation. After introducing a variable in such an intermediate representation, the position variable would probably needed to be introduced through some construction like Holland's local expectation value. Of all these theories, the one of \db\ and Bohm is arguably the simplest and most natural. It doesn't require a variable intermediate between the wavefunction and the position variable (like the momentum variable in Epstein's theory).

\section{Acknowledgments}
This work was initiated while at the Perimeter Institute for Theoretical Physics, Waterloo, Canada. Research there is supported by the Government of Canada through Industry Canada and by the Province of Ontario through the Ministry of Research \& Innovation. The work was further developed during an extended visit in the spring of 2008 at ISCAP, Columbia University, New York, USA. I am very grateful to David Albert, Brian Greene and Maulik Parikh for their invitation and discussions. Currently the support of the FWO-Flanders is acknowledged.

\end{document}